\documentclass[12pt]{iopart}
\usepackage{iopams}  
\usepackage{graphicx}
\begin{document}

\title
[Critical conductance and wavefunction of two dimensional Ando model]
{Critical regime of two dimensional Ando model: relation between critical conductance and
fractal dimension of electronic eigenstates}

\author{P. Marko\v{s}$^1$, L. Schweitzer$^2$}

\address{$^1$Institute of Physics, Slovak Academy of Sciences, D\'ubravsk\'a
  cesta 9, 854 11 Bratislava, Slovakia} 
\address{$^2$Physikalisch-Technische Bundesanstalt, Bundesallee 100, 38116
  Braunschweig, Germany} 
\eads{\mailto{peter.markos@savba.sk}, ~\mailto{ludwig.schweitzer@ptb.de}}
\begin{abstract}
The critical two-terminal conductance $g_c$ and the spatial fluctuations of
critical eigenstates 
are investigated for a disordered two dimensional model of non-interacting
electrons subject to spin-orbit scattering (Ando model). For square samples, 
we verify numerically the relation $\sigma_c=1/[2\pi(2-D(1))]\,e^2/h$ between 
critical conductivity $\sigma_c=g_c=(1.42\pm 0.005)\,e^2/h$ and the fractal 
information dimension of the electron wave function, $D(1)=1.889\pm 0.001$.  
Through a detailed numerical scaling analysis of the two-terminal conductance
we also estimate the critical exponent $\nu=2.80\pm 0.04$ that governs the
quantum phase transition.  
\end{abstract}

\maketitle

\section{Introduction}
Quantum phase transitions in two dimensional (2d) disordered systems have been the
subject of continuing research for many decades. In particular, systems of 
non-interacting particles that are invariant under time reversal, but lack the
symmetry of spin rotation are of special importance due to their property of a
complete Anderson transition \cite{HLN80,Mac84,Mac85,EZ87,Weg89,And89}. 
Thus, 2d systems with spin-orbit interaction,
which belong to the symplectic symmetry class \cite{ZP88}, 
exhibit an energy spectrum with regions of either localized or extended states
separated by quantum critical points at which the correlation length diverges
according to $\xi(E)\sim |E-E_c|^{-\nu}$ with a universal exponent $\nu$. 
The energetical position of these points $E_c$ depend on disorder strength and
spin-orbit interaction. The corresponding critical eigenstates were found to be
multi-fractal objects showing strong amplitude fluctuations as well as
long-range spatial and energetical correlations \cite{CDEN93,Sch95,SP99}.  

Our knowledge about the critical properties like critical energy and disorder,  
or about the exponents that govern their scaling, originate mainly
from numerical investigations of quantities like localization length
\cite{Mac85,EZ87,And89,Fea91,Eva95,ASO02} or spectral correlations of
eigenvalues \cite{SZ97} which both are not directly accessible to experimental
detection.  The aim of the present paper is to present scaling results of the
two-terminal  conductance \cite{ES81}, a physical quantity that can easily be
measured in experiment, and to examine a recently proposed relation between
the average critical conductance and the generalized fractal dimensions of
critical eigenstates. Therefore, we first present our numerical data for the
two-terminal conductance and obtain the critical parameters of the model by a
finite-size scaling analysis.
In the metallic regime, the mean conductance is expected to increase
with increasing system size while a decrease is usually found in the localized 
regime. At the critical point, the mean conductance $g_c$ should be a universal, 
system size independent quantity and equal to the critical conductivity
$\sigma$. Hence, the size scaling of our numerical data enables us to 
estimate the critical disorder $W_c$, and to calculate the critical 
conductance $g_c$. 
We also find the critical exponent $\nu$ which governs the energy and disorder
dependence of the conductance near the critical point.

Then we discuss the fractal properties of critical wave functions and
calculate fractal dimensions $D(q)$.    
Non-uniform spatial amplitude distributions of electron wave functions
that exhibit multi-fractal properties, are a characteristic feature of the critical
regime at metal-insulator transitions. It is well known that the corresponding 
generalized multi-fractal dimensions influence the time dependent transport
properties \cite{BHS96}. Here, we show that they also determine the 
conductance at a quantum critical point.
Our main result is a confirmation of the relation $D(q)=2-q/[\beta 4\pi^2 
\varrho D \hbar]$ between the fractal dimensions and the conductivity (per spin
direction), $\sigma=e^2 \varrho D$, proposed by Fal'ko and Efetov
\cite{FE95a}. We also examine the validity
of Janssen's formula \cite{Jan94,Jan98} which relates the
Lipschitz-H\"older exponent $\alpha_0$ of the maximum of the multi-fractal
distribution $f(\alpha)$ to the scaling parameter
$\Lambda_c=1/[\pi(\alpha_0-d)]$ 
for quasi-1d systems.
We find, however, our result for the symplectic Ando model ($d=2$) to be at
variance with this proposal. 

\section{The model and methods}

\subsection{Hamiltonian}

We study the two dimensional (2d) Ando model \cite{And89} defined on a square
lattice by the Hamiltonian 
\begin{equation}\label{hamiltonian}
{\cal H}=W\sum_n \varepsilon_nc^{\dag}_{n}c_{n}^{}+\sum_{[nn']}V_{nn'}c^{\dag}_{n}c_{n'}^{}.
\end{equation}
The $\varepsilon_n$ are uncorrelated random on-site energies chosen
according to a box probability distribution in the range $|\varepsilon_n|\le 1/2$. 
The parameter $W$ measures the strength of the disorder. 
The spin dependent hopping 
terms between nearest neighbour sites in the direction parallel and
perpendicular to the current flow are
\begin{equation}
V_\parallel=V\left(
\begin{array}{rr}
 c  & s\\
-s & c 
\end{array}
\right),
~~\textrm{and}~~~V_\perp=V\left(
\begin{array}{rr}
 c  & -is\\
-is & c 
\end{array}
\right),
\end{equation}
with $c^2+s^2=1$. 
The parameter $s$ determines hopping accompanied with a change of the
spin of the electron. As usual, it is set to $s=0.5$. The special case $s=0$
corresponds to a system with orthogonal symmetry which exhibits no
metal-insulator transition in 2d.  
Energies are measured in units $V=1$ and lengths
in units of the lattice constant $a=1$. 

It is well known that for Fermi energy $E_F\approx 0$, the Ando model exhibits
a metal-insulator transition as a function of disorder 
at $W=W_c\approx 5.8$
\cite{And89,Fea91}.
The critical exponent for various $2d$ systems with symplectic symmetry
has been estimated in numerous works using different numerical methods
\cite{Fea91,SZ97,MJH98,Min98,YO98,ASO02,ASO04},
leading to rather inconsistent results.  The critical conductance
was studied in \cite{OSK04}.  We will discuss some of these results later together
with our own results.

\subsection{Conductance}

The two-terminal conductance of a particular sample is calculated from
\begin{equation}\label{Landauer}
g=\Tr t^\dag t,
\end{equation}
where $t$ is the transmission matrix. Its elements $t_{\alpha\beta}$ determine the
transmission amplitude from channel $\alpha$ into channel $\beta$. 
Two semi-infinite ideal leads are attached to the disordered sample and
periodic boundary conditions are applied in the transverse direction.
The Fermi energy is fixed at $E_F=0.01$ and spin-orbit scattering is assumed
to be absent in the leads.

\begin{figure}[t!]
\begin{center}
\includegraphics[clip,width=0.65\textwidth]{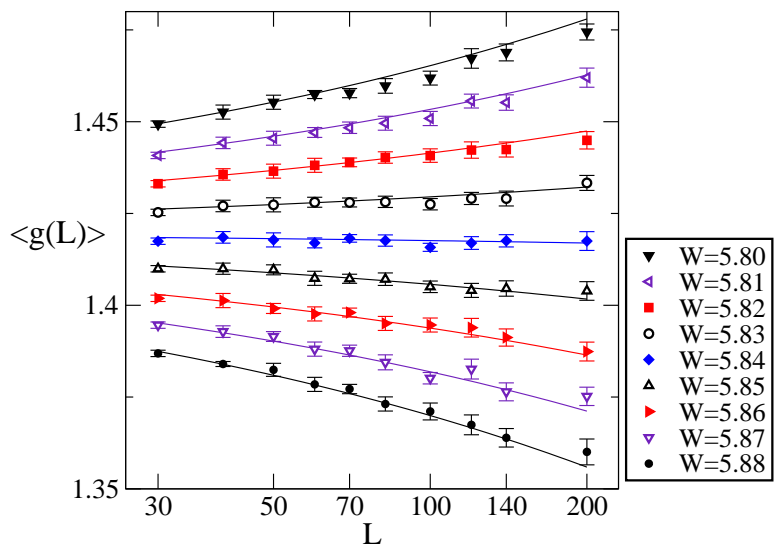}
\end{center}
\caption{The mean conductance $\langle g(L)\rangle$ as a function of the
  system size $L$ for 
  various values of disorder strength $W$. Within the interval of system sizes 
  $30\le L\le 200$, $\langle g(L)\rangle$ is almost constant for disorder
  $W=5.84$. This indicates that 
  finite size effects are absent and irrelevant scaling terms negligible.
  Solid lines are fits to \eref{scaling} with critical parameters
  $g_c=1.42$, $W_c=5.838  $, $\nu=2.80$ and $A=-0.23$. 
}
\label{fig1}
\end{figure}

We use the algorithm of Pendry \etal \cite{PMR92} for our numerical calculations.
The size of the sample varied from $L=20$ to $L=200$. 
We concentrate on the analysis of the $L$-dependence of the mean conductance 
in the neighbourhood of the critical point at $W_c\sim 5.80$. Because of the absence of
self-averaging of the conductance, we need to analyze data for a large number
of macroscopically identical samples which differ only by the respective microscopic
realization of the disorder.
For each disorder strength, $5.71\le W\le 5.96$, and each system size, we
collected a statistical ensemble of $N_{\rm stat}\ge 10^5$ 
samples and calculate the mean conductance $\langle g\rangle$ and variance 
$\textrm{var} g=\langle g^2\rangle -\langle g\rangle^2$. Here,
$\langle\dots\rangle$ means averaging over a statistical ensemble.
Mean values $\langle g\rangle$
with the uncertainty $\delta g=(\textrm{var} g/N_{\rm stat})^{1/2}$ are
used in the scaling analysis. 
As  a typical mean value $\langle g\rangle\approx 1.4$ and $\textrm{var} 
g \approx 0.36$, the relative uncertainty
$\delta g/\langle g\rangle$ of our data is of order of 0.0015.  Typical
numerical data are shown in \fref{fig1} and \ref{fig2}. 

\subsection{Critical wave function}
Electron wave functions at quantum critical points of disordered systems exhibit
a peculiar spatial structure \cite{Weg80a,Aok83}. For our symplectic model,
the critical wave functions for systems sizes of up to $L=260$ were obtained 
by direct diagonalization using a Lanczos-algorithm. The fractality of the
spatial fluctuations of the modulus of the normalized eigenstates $\psi_E(r)$ 
was determined from the power-law scaling of the $q$-th moment defined 
via a 'box-probability' 
\begin{equation}
P(q,\lambda)=\sum_i^{N(l)}(\sum_{r\in\Omega_i(l)}|\psi_E(r)|^2)^q
\sim \lambda^{\tau(q)}.
\label{boxprob}
\end{equation}  
The generalized fractal dimensions $D(q)=\tau(q)/(q-1)$ or
the so called $f(\alpha(q))$-distri\-bution
was thus derived  \cite{HP83,Hea86,CJ89},
where $\Omega_i(l)$ is the the i-th box of size $l=\lambda L$.
The $\tau(q)$ and $f(\alpha(q))$ are related by a Legendre transform.

\subsection{Scaling analysis}
\label{fss}
\begin{figure}[t]
\begin{center}
\includegraphics[clip,width=0.5\textwidth]{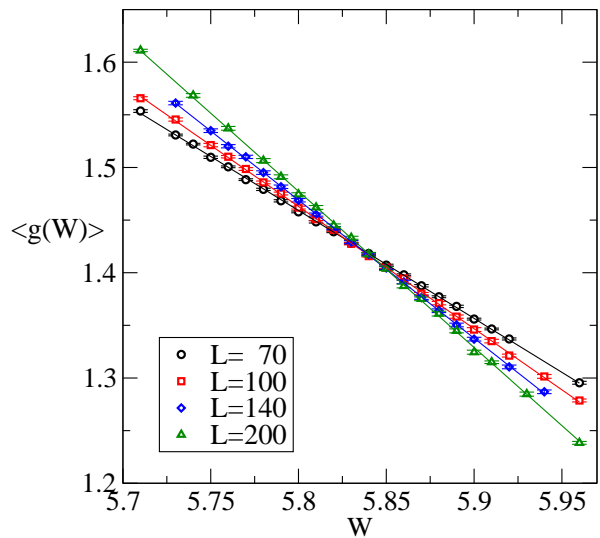}
\end{center}
\caption{Disorder dependence of the mean conductance $\langle g(W)\rangle$ for 
   $L=70$, 100, 140, and 200. Solid lines are linear fits \eref{linear}.  
}
\label{fig2}
\end{figure}

At the critical point $W\equiv W_c$, the conductance does not depend on
the system size. In accordance with the scaling theory of localization
\cite{MK81}, we assume that in the
neighbourhood of the critical point the $L$ dependence of $g$ is governed by
the critical exponent $\nu$,
\begin{equation}\label{scaling}
g_{\rm sc}(W,L)=g_c+A(W-W_c)L^{1/\nu}.
\end{equation}
We collected numerical data $g(W,L)$ for more than $N=200$ values of $W$ and $L$
and obtain the critical parameters, $g_c$, $W_c$ and $\nu$ from a fit of the
numerical data to the scaling ansatz (\ref{scaling}).
We made sure that more sophisticated fits \cite{SO99} which include higher
order terms in disorder are not needed because
our data exhibit already a perfect linear $W$-dependence (see \fref{fig2}),
Also, subject to the uncertainty of our raw data, we found that possible
irrelevant scaling fields \cite{SO99,Mac94} are weak and do only marginally
influence our results.  

The most simple analysis of critical parameters can be done as follows: first, 
we calculate parameters of the linear $W$-dependence of the conductance
\begin{equation}\label{linear}
g(W,L)=g_0(L)+Wg_1(L).
\end{equation}
Comparing \eref{linear} with \eref{scaling}, we see that the slope $g_1$ depends 
on the system size as
\begin{equation}\label{slope}
g_1(L)\propto L^{1/\nu}.
\end{equation}
Then, a power-law fit $g_1(L)$ \textit{vs.\@} $L$ gives us the critical exponent.

We also  minimize numerically the function
\begin{equation}\label{F}
F=\frac{1}{N}\sum_{LW} \frac{1}{(\delta g)^2}
\left[\langle g(W,L)\rangle-g_{\rm sc}(W,L)\right]^2
\end{equation}
with respect to unknown parameters $W_c$, $\nu$ and $g_c$ and $A$. To estimate
the accuracy of the final result, we repeat $N_m$ times the minimization of the
function $F$ with input data $\langle g(W,L)\rangle$ 
randomly fluctuating within their respective error bars. 
To understand the role of the finite size effects, we use in the scaling analysis
only data for systems with sizes $L_{\rm min}\le L\le L_{\rm max}$ and study
how the obtained critical parameters depend on the choice of $L_{\rm min}$ and
$L_{\rm max}$.

\begin{figure}[t!]
\begin{center}
\includegraphics[clip,width=0.5\textwidth]{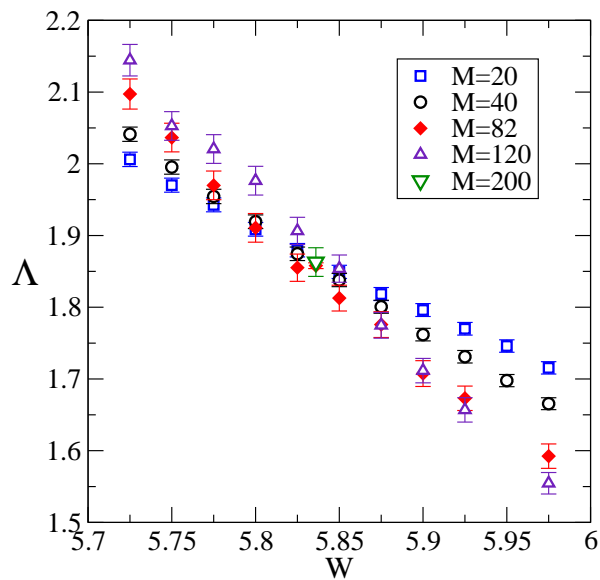}
\end{center}
\caption{%
The scaling parameter $\Lambda$ calculated for quasi-1d geometry versus
disorder strength for varying system width $M$. The data confirm a
critical disorder $W_c\approx 5.835$ and $\Lambda_c\simeq 1.87\pm 0.02$.
}
\label{q1d}
\end{figure}

\section{Results and discussions}

\subsection{Critical parameters}
Our analysis showed that neither the critical disorder nor the critical
conductance depend on the choice of the minimal and maximal system size,
$L_{\rm min}$ and $L_{\rm max}$.  From all the data we concluded that
\begin{equation}
W_c=5.838\pm 0.007
\end{equation}
The estimate of the critical disorder differs considerably from the result 
of Fastenrath \etal \cite{Fea91} ($W_c=5.74$), 
obtained from scaling analysis of the localization length of quasi-1d systems.
Since previous results were derived from small system's widths, 
the discrepancy could be explained as an effect of finite size corrections. 
To support this assumption, we performed numerical simulations  
for quasi-1d systems of size $M\times L$.  Figure \ref{q1d} shows the
calculated data for the ratio $\Lambda=\lambda_M/M$,
the localization length $\lambda_M$ divided by the system width $M$
\cite{MK81,PS81}.  For our purposes, we run $\Lambda$ 
with an accuracy of only 1\%. This was sufficient to confirm 
that our estimation of the critical disorder derived from the conductance
calculations indeed coincides with the one obtained from the localization length. 
Also, we get an estimate of the critical value $\Lambda_c$
of the scaling parameter $\Lambda$, 
\begin{equation}\label{lambdac}
\Lambda_c\simeq 1.87\pm 0.02
\end{equation}
which agrees well with a recent result of Asada \etal \cite{ASO04}
obtained for a SU(2) model.

\begin{figure}[t!]
\begin{center}
\includegraphics[clip,width=0.5\textwidth]{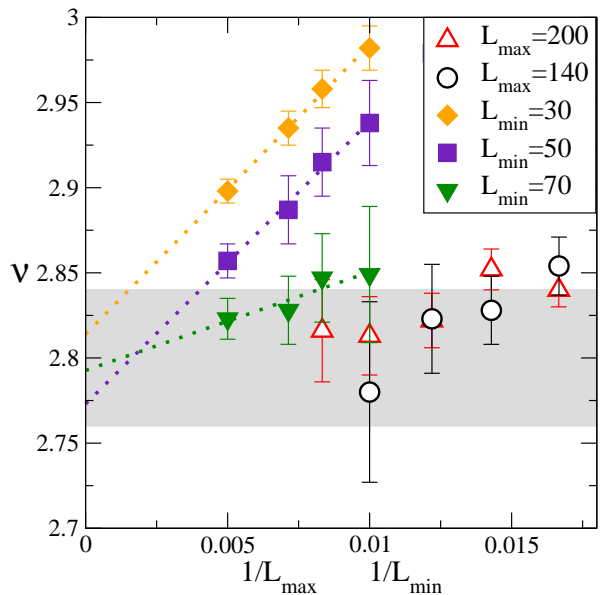}
\end{center}
\caption{%
The critical exponent $\nu$ obtained from numerical data
with $L_{\rm min}\le L\le L_{\rm max}$. Full symbols 
shows how $\nu$ depends on the choice of $L_{\rm max}$ for
fixed minimal system size $L_{\rm min}$. Open symbols show
how $\nu$ depends on $L_{\rm min}$ for fixed $L_{\rm max}$,
$L_{\rm max}=140$ ({\large$\circ$}),
and $L_{\rm max}=200$ ($\bigtriangleup$).   The shaded area 
highlights our estimate of the critical exponent $\nu=2.80\pm 0.04$.
}
\label{fig3}
\end{figure}

From our conductance data, the critical exponent $\nu$ was obtained for the first
time using finite size scaling of the calculated electrical two-terminal conductance, 
which is, unlike the localization length, an easy way to measure physical quantity.
With the finite size scaling analysis described in section~(\ref{fss}), our
estimation of the critical exponent gives 
\begin{equation}\label{nu}
\nu=2.8\pm 0.04,
\end{equation}
which is close to a numerical result $\nu=2.746\pm 0.009$ for a SU(2) model
published recently \cite{ASO04}, but considerably larger than $\nu=2.05\pm 0.08$
\cite{And89} and smaller than $\nu=2.88\pm 0.15$ as obtained in \cite{Min98}.  
In contrast to the critical conductance and critical disorder, the critical 
exponent is sensitive to the system size as can be seen in \fref{fig3}.
The estimate of $\nu$ is therefore highly non-trivial and error bars obtained from 
various fit procedures differ. 

\subsection{Critical conductance}
The critical two-terminal conductance obtained by us for square samples with
periodic boundary conditions in the transverse direction,
\begin{equation}\label{gc}
g_c=1.42\pm 0.005,
\end{equation}
can be compared with the result for a SU(2) model studied recently in \cite{OSK04}.
From the corresponding numerical data for size $L\le 97$ (T. Ohtsuki, private
communication) we calculated $\lim_{L\to\infty}g(L)=1.411$, which is in
very good agreement with our present result. 

Assuming for square systems $\sigma \equiv g/L^{(2-d)}$, we have at the
critical point  $g_c=\sigma_c$.  As mentioned above, for $L>20$ our critical
conductance data do not exhibit any finite size effect. We do not expect that
the numerical analysis causes any inaccuracy of the estimation of $g_c$.
Still, when comparing $g_c$ with  $\sigma_c$, we have to
keep in mind that small corrections due to the properties of leads (we assume
perfect leads without spin-orbit scattering) might be responsible for a small
difference between $\sigma_c$ and $g_c$. However, we do not expect this
difference to be larger than $\sim L^{-1}$ and neglect it for further purposes.

\begin{figure}[t!]
\begin{center}
\includegraphics[clip,width=0.5\textwidth]{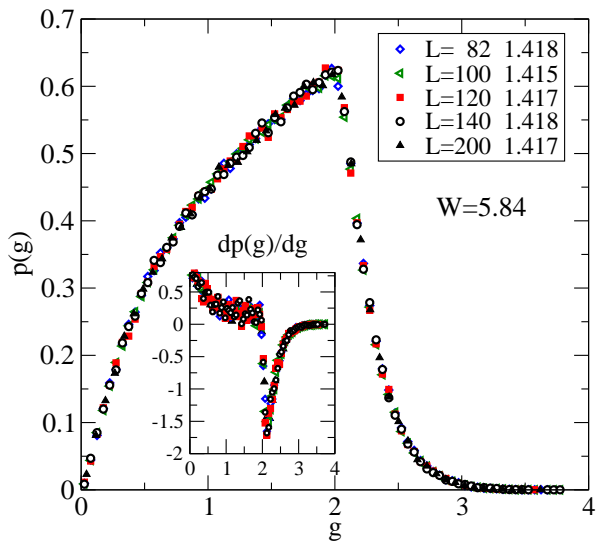}
\end{center}
\caption{Critical conductance distribution $p(g)$.  
Inset: the derivative $\partial p(g)/\partial g$ which shows $g_{\rm
  non-an}$ to be slightly larger than 2. 
}
\label{fig4}
\end{figure}

\Fref{fig4} shows the critical conductance distribution $p(g)$. We present 
data for disorder $W=5.84$ and system size $82\le L \le 200$.  
We see that indeed $p(g)$ does not depend on the system size at the
critical point.
Besides the 
shape of the distribution, which is known also from previous studies, 
we want to point out the non-analytical behaviour 
of the distribution at the point $g=g_{\rm non-an}$ which 
is slightly larger than 2. This agrees with a qualitative 
estimation of Muttalib \etal \cite{MWGG03},
although their work concerns quasi one dimensional 
systems with orthogonal symmetry.

\subsection{Fractal dimensions and Fal'ko-Efetov relation}

\begin{figure}[t]
\begin{center}
\includegraphics[clip,width=0.55\textwidth]{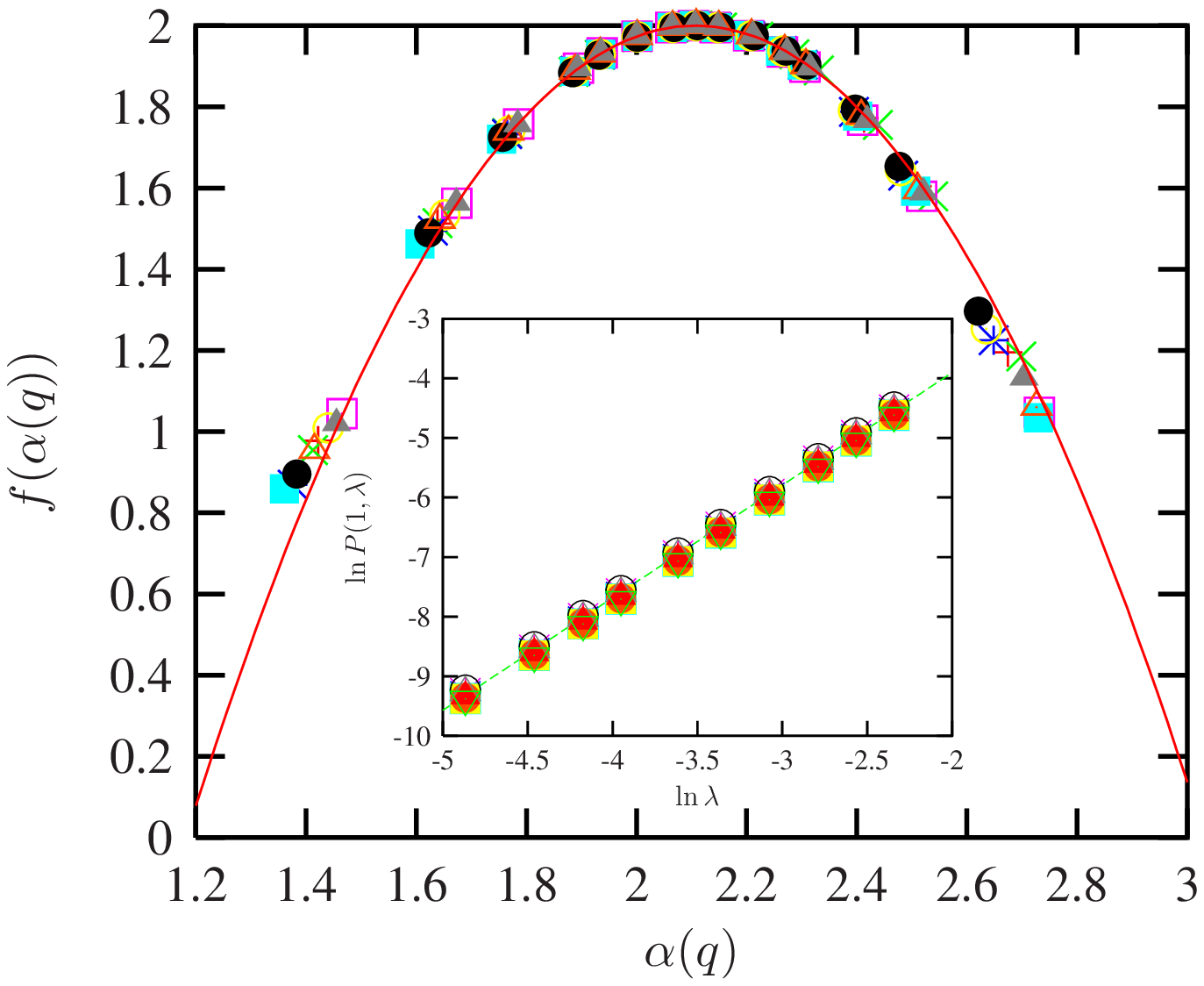}
\end{center}
\caption{%
The $f(\alpha(q))$-distributions of 10 critical eigenstates at $E\approx 0$ 
and $W_c=5.83$ for $q=0.0,\pm 0.2, \pm 0.5, \pm 0.8, \pm 1.0, \pm 1.5, \pm
2.0$, and $\pm 3.0$. The parabolic fit is determined by 
$\alpha_0=2.107\pm 5\cdot 10^{-3}$. 
The inset shows the scaling of $\ln P(1,\lambda)$ vs.\@
logarithm of the box size $\lambda$. Data are obtained from ten $L=260$
samples with different disorder realization. The dashed line has 
steepness $\tau(1)=1.889$.
}
\label{D1scal}
\end{figure}

In figure~\ref{D1scal} the $f(\alpha(q))$-distributions of 10 critical
eigenfunctions of systems with $L=260$ are shown together with a parabolic fit 
$f(\alpha(q))=d-(\alpha(q)-\alpha_0)^2/[4(\alpha_0-d)]$ 
with $\alpha_0=\alpha(q=0)=2.107\pm 0.005$. This value is lesser than
the one obtained previously for smaller sample size $L=150$ \cite{Sch95}. 
The inset exhibits the logarithm of $P(1,\lambda)$ plotted versus $\ln
\lambda$ for $2\le l \le 25$. The data belonging to critical eigenstates 
obtained from 10 disorder realizations clearly obey power-law scaling.

From the multi-fractal analysis of critical eigenfunctions of $L=260$ samples,
we obtained an information dimension $D(1)=1.889\pm 0.001$
using 10 different realizations with disorder strength $W=5.83$.
With  this data, we checked the validity of a formula suggested by Fal'ko and
Efetov \cite{FE95a} which should hold in the limit $1\le \varrho D h$,
\begin{equation}\label{sigma-dq}
D(q) = 2-\frac{q}{\beta 4\pi^2\varrho D \hbar},
\label{FalEfe}
\end{equation}
where $\beta=1/2,~1$ and 2 for or orthogonal, unitary and symplectic symmetry, respectively.
With $\sigma=e^2\varrho D$ (diffusion constant $D$ and density
of states $\varrho$) we get
\begin{equation}\label{efetov}
\sigma=\displaystyle{\frac{q}{2\pi\beta[2-D(q)]}\frac{e^2}{h}}.
\end{equation}

\begin{figure}[t]
\begin{center}
  \includegraphics[clip,width=0.5\textwidth]{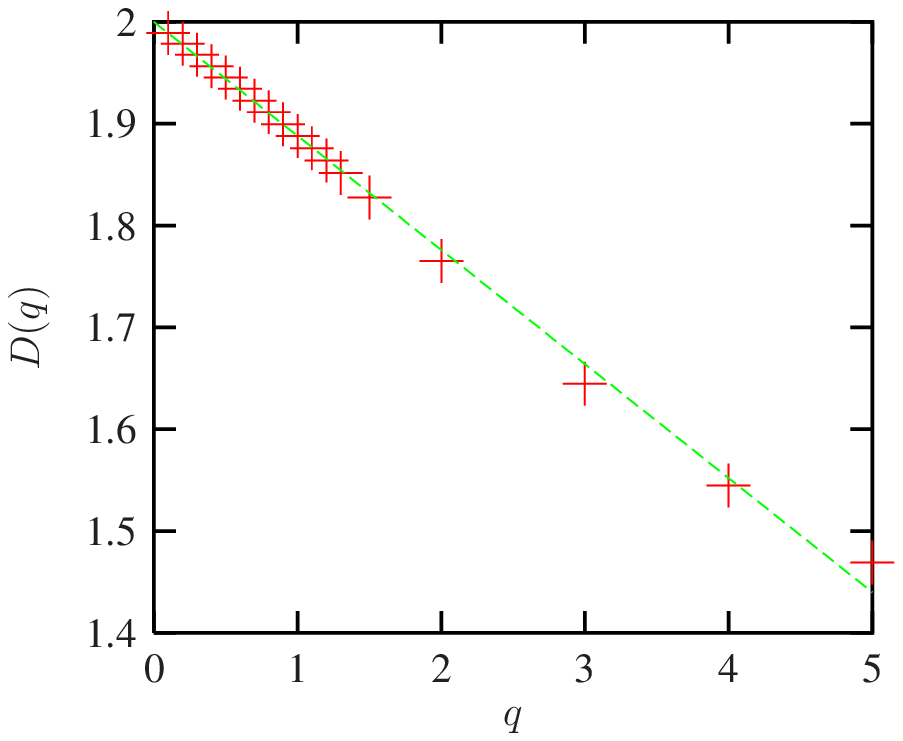}
\end{center}
\caption{%
The generalized fractal dimension $D(q)$ of a critical eigenfunction as a 
function of $q$ calculated via the box-probability (\ref{boxprob}). The
linearity $D(q)=2-q/k$ holds for $q\lesssim 1.5$, with $k=0.112$ leading to
$\sigma_c=0.71\,e^2/h$.  
}
\label{Dq}
\end{figure}

Taking $q=1$, we found our results for $\sigma_c$ and $D(1)$ to be in very
good agreement with (\ref{efetov}).  
Therefore, this relation is fulfilled also for the symplectic symmetry in a similar
manner as found previously for the quantum Hall case \cite{SM05}. In comparing
these results one has to keep in mind that in the present paper two spin
channels were considered, i.e., the conductance has to be divided by 2, whereas
in the QHE case only one spin direction was taken into account. 
Note also that the value of the disorder used in the multi-fractal 
analysis, $W=5.83$, differs slightly form the critical disorder
$W_c=5.838$. Therefore, the conductance $g(W=5.83)\approx 1.43$ mentioned 
in ref.~\cite{SM05} is a little larger than $g_c$.
The validity of the proposed linear relationship (\ref{FalEfe}) holds at least
for $q \le 1.5$ which can be seen in figure~\ref{Dq} where $D(q)$ calculated
according to (\ref{boxprob}) is plotted versus $q$.

\subsection{Comparison with Janssen's formula}
A formula that connects fractal properties of critical eigenstates in
square samples with the finite size scaling variable $\Lambda=\lambda_M/M$ of
quasi-1d systems was conjectured by Janssen \cite{Jan94,Jan98}. At the
critical point, $\Lambda_c$ is scale independent and depends only on the
Lipschitz-H\"older exponent $\alpha_0$ and the spatial dimension of the
system $d=2$ 
\begin{equation}\label{Lambda}
\Lambda_c=\displaystyle{\frac{1}{\pi(\alpha_0-d)}}.
\end{equation}
Equation (\ref{Lambda}) was reported to hold for 2d disordered systems in the quantum
Hall regime \cite{Jan94,Huc94}. In the present symplectic system, however, it is not
met by our results. 
Also, assuming the parabolic approximation to be valid, we can use the relation
between $D(1)$ and $\alpha_0$,  
\begin{equation}
D(1)=\alpha(1)=4-\alpha_0,
\end{equation}
and connect (\ref{efetov}) and (\ref{Lambda}) to obtain a
relation between $\Lambda_c$ and the critical conductivity 
\begin{equation}\label{univ}
\sigma_c=\frac{\Lambda_c}{2\beta}\frac{e^2}{h}.
\end{equation}
Again, contrary to what is observed in the QHE regime, the relation
(\ref{univ})  
is not satisfied for a symplectic system described by the Ando model.
A possible reason for this failure might be that in the present 2d model, the
scaling parameter $\Lambda_c$ may not represent the typical localization
length of the system as has been assumed in \cite{Jan94}.

\section{Conclusion}

We studied the electrical two-terminal conductance and the spatial
fluctuations of electron eigenfunctions near the metal-insulator transition
of the two dimensional Ando model. Using finite-size scaling, we obtained for 
this symplectic model a critical exponent $\nu=2.8\pm 0.04$, which governs
the size dependence of the conductance in the critical regime,
and a critical disorder $W_c=5.838\pm 0.007$.
Our results for the critical conductance and for the fractal dimension of
critical eigenstates confirm the validity of Fal'ko and Efetov's prediction
(\ref{sigma-dq}). However, comparison of the fractal dimensions with the
critical value of the finite size scaling parameter $\Lambda_c$, calculated
for quasi one-dimensional systems, indicates that Janssen's formula
(\ref{Lambda}) is not fulfilled by our data in the present model. 

Unlike the localization
length, the two-terminal conductance should be easily accessible in experiments.
Our numerical confirmation of the relation between fractal dimensions
of critical eigenfunctions and the critical conductance provides an
additional argument for universality at the metal-insulator transition
in 2d symplectic models.

\ack
We thank T. Ohtsuki for sending us numerical data of the
critical conductance from the SU(2) model.
PM thanks APVT Grant No. 51-021602 for partial financial support, and LS
for the hospitality of the Slovak Academy of Sciences.

\section*{References}
\providecommand{\newblock}{}

\end{document}